# CONSTRAINED DYNAMICS OF TACHYON FIELD IN FRWL SPACETIME[†]

*UDC 531.17, 524.83*

**Marko A. Dimitrijević[1], Dragoljub D. Dimitrijević[2], Goran Đorđević[2], Milan Milošević[2]**

[1]Faculty of Electronic Engineering, University of Niš,
Aleksandra Medvedeva 14 Street, 18000 Niš, Serbia
[2]Department of Physics, Faculty of Science and Mathematics, University of Niš,
Višegradska 33 Street, 18000 Niš, Serbia

**Abstract**. *In this paper, we continue the study of tachyon scalar field described by a Dirac-Born-Infeld (DBI) type action with constraints in the cosmological context. The proposed extension of the system introducing an auxiliary field in the minisuperspace framework is discussed. A new equivalent set of constraints is constructed, satisfying the usual regularity conditions.*

**Key words**: *tachyon field, cosmology, constrained system, Dirac-Bergman algorithm*

1. INTRODUCTION

The understanding of constrained dynamical systems requires knowledge of classical Hamiltonian theory. The study of a system with constraints usually starts with the Hamiltonian formalism and the Dirac-Bergman algorithm (Bergmann, 1949; Dirac, 1950). This algorithm provides a scheme to classify constraints into the first and the second class.

The presence of constraints in the singular theories requires a careful analysis when applying the Dirac-Bergman algorithm. Dirac showed that the algebra of Poisson's brackets divides the constraints into the first- and the second-class constraints (Dirac, 1964). The first-class constraints have zero Poisson's brackets with all other constraints

Received March 29th, 2019; accepted April 25th, 2019
**Corresponding author**: Marko A. Dimitrijević
Faculty of Electronic Engineering, University of Niš, Aleksandra Medvedeva 14 Street, 18000 Niš, Serbia
E-mail: marko@venus.elfak.ni.ac.rs
[†] This work has been supported by the ICTP - SEENET-MTP project NT-03 Cosmology-Classical and Quantum Challenges. The financial support of the Serbian Ministry for Education, Science and Technological Development under the project OI 174020 and OI 176021 is also kindly acknowledged.

(in the subspace of the phase space in which constraints hold), while constraints that do not belong to the first-class set, by definition, belong to the second-class constraints.

In this paper, we examine tachyon scalar field minimally coupled with gravity in the Friedmann-Robertson-Walker-Lemaitre (FRWL) background in the minisuperspace context. This model is a starting point in order to understand the potential role of models with a non-canonical Lagrangians in cosmology. As an example, we consider the DBI type Lagrangian $\mathcal{L}_{tach}[T]$ for the tachyon scalar field $T$ (Sen, 1999; Steer and Vernizzi, 2004)

$$\mathcal{L}_{tach}[T] = -V(T)\sqrt{1 + g^{\mu\nu}\partial_\mu T \partial_\nu T}, \tag{1}$$

where $V(T)$ is the tachyon field potential and $g_{\mu\nu}$ denotes components of the metric tensor with $(-,+,+,+)$ signature. The full equation of motion is of the form

$$\left(g^{\mu\nu} - \frac{\partial^\mu T \partial^\nu T}{1-(\partial T)^2}\right)\partial_\mu \partial_\nu T = -\frac{1}{V(T)}\frac{dV}{dT}(1-(\partial T)^2). \tag{2}$$

In the cosmological context, scalar fields are used to describe the perfect cosmological fluid with the pressure $P$ and the matter density $\rho$

$$P(\partial T, T) \equiv \mathcal{L}[T], \tag{3}$$

$$\rho(\partial T, T) \equiv 2X\frac{\partial \mathcal{L}}{\partial X} - \mathcal{L}[T], \tag{4}$$

where

$$X(\partial T) = \frac{1}{2} g^{\mu\nu}\partial_\mu T \partial_\nu T. \tag{5}$$

Even in the case of the spatially homogenous tachyon scalar field (i.e. minisuperspace model where all quantities depend on time coordinate only) the model already contains (the first class) constraints. This is due to the reparametrization invariance of the pure gravitational part. Extending the phase space of the model in order to avoid appearances of the square root by introducing an auxiliary field and its conjugate momentum, one can get the modified action (Dimitrijevic et al., 2019). The modified action will contain additional constraints. This is explained in what follows and it is the main subject of this paper.

2. A NON-STANDARD LAGRANGIAN AND THE MINISUPERSPACE MODEL

To describe the cosmological space-time background in the minisuperspace framework, we will restrict on homogeneous and isotropic space, taking the FRWL four-dimensional space-time metric (in natural units, $c = 1$):

$$ds^2 = -N^2(t)dt^2 + a^2(t)d\vec{x}^2, \tag{6}$$

where $a(t)$ denotes the scale factor, and $N(t)$ is the lapse function. The total action is the sum of the term which describes gravity (Ricci scalar curvature, $R$) and the term that describes cosmological fluid through the minimally coupled spatially homogenous tachyon scalar field $T$, with the Lagrangian $\mathcal{L}_{tach}[T]$

$$S = -\frac{1}{16\pi G}\int d^4x\sqrt{-g}R + \int d^4x\sqrt{-g}\mathcal{L}_{tach}[T], \qquad (7)$$

where $g$ is the determinant of the metric tensor for the FRWL space-time. In the minisuperspace framework we have (using $16\pi G = 1$)

$$S^{(0)} = \int dt \mathcal{L}, \qquad (8)$$

where the tachyon scalar field part of the total Lagrangian $\mathcal{L}$, written as

$$\sqrt{-g}\mathcal{L}_{tach}[T,\dot{T},N,a] = -Na^3 V(T)\sqrt{1 - N^{-2}\dot{T}^2}, \qquad (9)$$

is extended to the classically equivalent one

$$\sqrt{-g}\mathcal{L}_{ext}[T,\dot{T},N,a,\sigma] = Na^3\left(\frac{1}{2N^2}\frac{\dot{T}^2}{\sigma} - \frac{1}{2}\sigma V^2(T) - \frac{1}{2\sigma}\right). \qquad (10)$$

Here, the dot represents derivation w.r.t. time coordinate *t*. Details concerning this extension, i.e. how to introduce an auxiliary scalar field $\sigma$ and to obtain the total Lagrangian $\mathcal{L}$, can be found in Dimitrijevic et al. (2019). The final expression, which will be our starting point, is

$$\mathcal{L} = -6\frac{a\dot{a}^2}{N} + 6kNa + \frac{a^3}{2N}\frac{\dot{T}^2}{\sigma} - \frac{1}{2}\sigma Na^3 V^2(T) - \frac{Na^3}{2\sigma}, \qquad (11)$$

where $k$ is a spatial curvature. It is obvious that the model will have two primary constraints, i.e. two momenta conjugate to the auxiliary field $\sigma$ and the lapse field $N$.

## 3. IMPORTANT MATRICES

In this section, we recapitulate some important matrices that are used when a singular system is studied. Most of the definitions follow the book by Gitman and Tyutin (Gitman and Tyutin, 1990).

If a theory is a singular one, then among its equations of motion in the (extended) Hamiltonian formalism, some constraints exist obligatory.

The first important matrix is the Hessian matrix $\hat{H}$, with the rank $R_H$. It is a quadratic matrix defined as

$$\hat{H} = \left[\frac{\partial^2 \mathcal{L}}{\partial \dot{q}_i \partial \dot{q}_j}\right], \quad \text{rank } \hat{H} = R_H, \qquad (12)$$

where $q_i$ are the configuration variables (or fields). If the Hessian is a singular matrix, $\det \hat{H} = 0$, then the corresponding system is called singular, i.e. it contains primary constraints. The number $\alpha$ of primary constraints is

$$\alpha = N - R_H, \tag{13}$$

where $N$ is the number of configuration variables.

By virtue of the Dirac-Bergman algorithm, the consistency conditions for the primary constraints lead to the secondary constraints, while the consistency conditions for the secondary constraints lead to the tertiary constraints, etc. After this algorithm is terminated, we end up with the total number $n$ of all constraints.

The set of primary constraints is not always functionally independent. It means that the number of constraints in the true set of primary constraints could differ from $\alpha$. In regard to this, the number $n_1$ of functionally independent primary constraints $\Phi_i^{(1)}$ is equal to the rank $R_{\hat{A}^{(1)}}$ of (rectangular) matrix $\hat{A}^{(1)}$:

$$\hat{A}^{(1)} = \left[\frac{\partial \Phi_i^{(1)}}{\partial (q_j, p_k)}\right], \quad \text{rank } \hat{A}^{(1)} = R_{\hat{A}^{(1)}}, \tag{14}$$

where $p_k$ are the momenta conjugate to the configuration variables. Regularity condition means that equality

$$n_1 \equiv R_{\hat{A}^{(1)}} = n - R_H, \tag{15}$$

is satisfied, where, again, $n$ is the number of all (primary, secondary…) constraints.

The set of all constraints $\Phi_i$ is functionally independent if their number $n$ is equal to the rank $R_A$ of a (rectangular) matrix $\hat{A}$:

$$n \equiv R_A, \tag{16}$$

$$\hat{A} = \left[\frac{\partial \Phi_i}{\partial (q_j, p_k)}\right], \quad \text{rank } \hat{A} = R_A. \tag{17}$$

The next two important matrices are defined through the Poisson bracket. If the rank $R_{\hat{M}^{(1)}}$ of a rectangular matrix $\hat{M}^{(1)}$ defined as

$$\hat{M}^{(1)} = \left[\{\Phi_i, \Phi_j^{(1)}\}_{PB}\right], \quad \text{rank } \hat{M}^{(1)} = R_{\hat{M}^{(1)}}, \tag{18}$$

is not maximal, then the system contains $\mu_1$ independent linear combinations of primary constraints, which are first-class quantities, where $\mu_1$ is defined as

$$\mu_1 = n_1 - R_{\hat{M}^{(1)}}\big|_{\Phi_i = 0}. \tag{19}$$

In a similar way, an antisymmetric matrix $\hat{M}$ is composed of the Poisson brackets of all constraints

$$\hat{M} = \left[\{\Phi_i, \Phi_j\}_{PB}\right], \quad \text{rank } \hat{M} = R_M. \tag{20}$$

If the matrix $\hat{M}$ is a singular one, then the corresponding system contains $\mu$ independent linear combination of constraints, which are first-class quantities, where $\mu$ is defined as

$$\mu = n - R_M\big|_{\Phi_i = 0}. \tag{21}$$

These important numbers help us to count and classify various types of the constraints (see Table 1). Note that $\mu_1$ and $\mu$ are defined on the constraint hypersurface and the number of all second class constraints needs to be even.

**Table 1** Important numbers defined through the rank of important matrices

| Symbol | Number of constraints |
|---|---|
| $\alpha$ | primary |
| $n_1$ | true primary |
| $n$ | all |
| $\mu_1$ | first-class primary |
| $\mu$ | all first class |
| $n - \mu$ | all second class |

A very important fact, which will be used in what follows, is that for any set of constraints $\Phi_i$ there exists an equivalent set of constraints $\Psi_i$ (Henneaux and Teitelboim, 1992). The new, equivalent set will consist of the first-class and the second-class constraints, and the number of constraints of a new set will be equal to the number of constraints of the starting set.

## 4. COUNTING AND CLASSIFICATION OF THE CONSTRAINTS

The starting set of constraints for the system with the Lagrangian (11) is known (Dimitrijevic et al., 2019) and consists of two primary and two secondary constraints. The primary constraints are written as

$$\Phi_1 = p_\sigma \approx 0, \tag{22}$$

$$\Phi_2 = p_N \approx 0, \tag{23}$$

while the secondary constraints arising from the consistency condition of the primary constraints

$$\dot{\Phi}_1 = \{\Phi_1, \mathcal{H}_T\}_{PB} \approx 0, \tag{24}$$

$$\dot{\Phi}_2 = \{\Phi_2, \mathcal{H}_T\}_{PB} \approx 0, \tag{25}$$

are

$$\Phi_3 = \Omega \approx 0, \tag{26}$$

$$\Phi_4 = \mathcal{H}_0 \approx 0. \tag{27}$$

The fundamental Poisson brackets are defined in the usual way

$$\{q_i, p_j\}_{PB} = \delta_{ij}, \tag{28}$$

The consistency conditions for the secondary constraints give no new constraints. The total Hamiltonian is

$$\mathcal{H}_T \equiv \mathcal{H}_c + \lambda_1 \Phi_1 + \lambda_2 \Phi_2 = N\mathcal{H}_0 + \lambda_1 \Phi_1 + \lambda_2 \Phi_2, \tag{29}$$

where $\lambda_1$ and $\lambda_2$ are Lagrange multipliers, and

$$\Omega \equiv \frac{1}{2a^3} p_T^2 + \frac{1}{2} a^3 V^2(T) - \frac{a^3}{2\sigma^2} , \qquad (30)$$

$$\mathcal{H}_0 \equiv -\frac{1}{24} \frac{p_a^2}{a} - 6ka + \frac{\sigma}{2a^3} p_T^2 + \frac{1}{2} \sigma a^3 V^2(T) + \frac{a^3}{2\sigma} . \qquad (31)$$

The constraints $\Phi_1$, $\Phi_3$ and $\Phi_4$ are of the second class, which is impossible (as already said, this number has to be even). The important numbers for the set of constraints $\Phi_i$ are given in Table 2.

**Table 2** Important numbers for the system with the Lagrangian (11) and the set of constraints $\Phi_i$

| Symbol | Number of constraints |
|---|---|
| $\alpha = 2$ | primary |
| $n_1 = 2$ | true primary |
| $n = 4$ | all |
| $\mu_1 = 1$ | first-class primary |
| $\mu = 1$ | all first class |
| $n - \mu = 3$ | all second class |

From the last row in Table 2 we see that the starting set of constraints $\Phi_i$ contains an odd number of second-class constraints, which means the constraints are not functionally independent. We need to find an equivalent set of constraints $\Psi_i$, which will have an appropriate i.e. even number of the second-class constraints. Note the FRWL minisuperspace model for the empty Universe contains two first-class constraints (Halliwell, 1988), while our calculations signalize that the extended Lagrangian (11) possesses two second-class constraints. Hence, we will construct a new set which will consist of the two first-class constraints ($\mu = 2$) and the two second-class constraints.

## 5. EQUIVALENT SET OF CONSTRAINTS

The starting point of the procedure of constructing an equivalent set of constraints $\Psi_i$ is a homogenous set of algebraic equations

$$\{\Phi_i, \Phi_j\}_{PB} u^j_{(\xi)} \approx 0, \quad \xi = 1, \ldots, \mu', \quad i, j = 1, \ldots, n , \qquad (32)$$

where $u^j_{(\xi)}$ are some unknown objects (in general, they are functions of the phase-space coordinates) that need to be found. Note that we assume here the number $\mu'$ of all first-class constraints to be equal to two, as discussed at the end of the previous section, which will be proved as an accurate assumption.

The set (32) for $n = 4$ becomes

$$-\frac{\partial \Omega}{\partial \sigma} u_{(\xi)}^3 \approx 0, \tag{33}$$

$$0 = 0, \tag{34}$$

$$\frac{\partial \Omega}{\partial \sigma} u_{(\xi)}^1 + \{\Omega, \mathcal{H}_0\}_{PB} u_{(\xi)}^4 \approx 0, \tag{35}$$

$$\{\mathcal{H}_0, \Omega\}_{PB} u_{(\xi)}^3 \approx 0, \tag{36}$$

with the general set of solutions

$$u_{(\xi)}^1 = \left(\frac{\partial \Omega}{\partial \sigma}\right)^{-1} \{\mathcal{H}_0, \Omega\}_{PB} u_{(\xi)}^4, \quad u_{(\xi)}^3 = 0, \tag{37}$$

$$u_{(\xi)}^2, u_{(\xi)}^4 - \text{arbitrary}. \tag{38}$$

Now, we can choose $u_{(1)}^4 = 0$, $u_{(1)}^2 = 1$ for (for $\xi = 1$) and $u_{(2)}^4 = 1$, $u_{(2)}^2 = 0$ (for $\xi = 2$), obtaining

$$\hat{u}_{(1)} = \begin{bmatrix} 0 \\ 1 \\ 0 \\ 0 \end{bmatrix}, \quad \hat{u}_{(2)} = \begin{bmatrix} \left(\frac{\partial \Omega}{\partial \sigma}\right)^{-1} \{\mathcal{H}_0, \Omega\}_{PB} \\ 0 \\ 0 \\ 1 \end{bmatrix}. \tag{39}$$

With the help of the already obtained $u_{(\xi)}^j$, the next step is to construct a new set of constraints, starting with the first-class pair

$$\Psi_\xi = u_{(\xi)}^i \Phi_i, \quad \xi = 1, 2, \tag{40}$$

that leads to

$$\Psi_1 = p_N, \tag{41}$$

$$\Psi_2 = \left(\frac{\partial \Omega}{\partial \sigma}\right)^{-1} \{\mathcal{H}_0, \Omega\}_{PB} p_\sigma + \mathcal{H}_0. \tag{42}$$

The Poisson brackets of $\Psi_1$ and $\Psi_2$ between themselves and with all the constraints $\Phi_i$ vanish on the constraint surface, which confirms the fact they are indeed the first-class constraints.

To construct the second-class pair of constraints, we begin with a linear combination

$$\Psi_j = u_{(j)}^i \Phi_i, \quad i, j = 1, \ldots, n, \tag{43}$$

where $u_{(j)}^i$ are components of a matrix $\hat{u}$, which need to be nonsingular. Note that the first two columns of $\hat{u}$ are already defined with the column-matrix $\hat{u}_{(1)}$ and $\hat{u}_{(2)}$. The undefined additional $u_{(j)}^i$ components for the case $j = 3$ and $j = 4$ can be chosen to be

$$\hat{u}_{(3)} = \begin{bmatrix} 1 \\ 0 \\ 0 \\ 0 \end{bmatrix}, \quad \hat{u}_{(4)} = \begin{bmatrix} 0 \\ 0 \\ 1 \\ 0 \end{bmatrix}, \tag{44}$$

which means that $\hat{u}$ remains nonsingular (more precisely, $\det \hat{u} = -1$). In this way, the matrix $\hat{u}$ receives the form

$$\hat{u} = \begin{bmatrix} 0 & \left(\frac{\partial \Omega}{\partial \sigma}\right)^{-1} \{\mathcal{H}_0, \Omega\}_{PB} & 1 & 0 \\ 1 & 0 & 0 & 0 \\ 0 & 0 & 0 & 1 \\ 0 & 1 & 0 & 0 \end{bmatrix}. \tag{45}$$

Finally, we get the equivalent set of constraints, with a pair of the first-class ($\Psi_1$, $\Psi_2$) and a pair of the second-class ($\Psi_3$, $\Psi_4$) constraints:

$$\Psi_1 = p_N, \tag{46}$$

$$\Psi_2 = \left(\frac{\partial \Omega}{\partial \sigma}\right)^{-1} \{\mathcal{H}_0, \Omega\}_{PB} p_\sigma + \mathcal{H}_0, \tag{47}$$

$$\Psi_3 = p_\sigma, \tag{48}$$

$$\Psi_4 = \Omega. \tag{49}$$

The only nonvanishing Poisson bracket is

$$\{\Psi_3, \Psi_4\}_{PB} = -\frac{\partial \Omega}{\partial \sigma} \neq 0, \tag{50}$$

as expected. In this way, the obtained set has the appropriate, i.e. even number of the second-class constraints.

## 6. DISCUSSION AND CONCLUSION

The obtained new set of the constraints (46)-(49) will be useful in the quantization procedure, whether it be canonical quantization, with the use of Dirac brackets, or Feynman path integral procedure, where appropriate path integral measure needs to be defined. This is not a unique choice of the set of constraints, but it seems to be the most natural. Table 3 contains important numbers for the new set.

**Table 3** Important numbers for the system with the Lagrangian (11) and the new equivalent set of constraints $\Psi_i$

| Symbol | Number of constraints |
|---|---|
| $\alpha = 2$ | primary |
| $n_1 = 2$ | true primary |
| $n = 4$ | All |
| $\mu_1 = 1$ | first-class primary |
| $\mu' = 2$ | all first class |
| $n - \mu' = 2$ | all second class |

It is worth mentioning that the extended model for the tachyon field (11) without gravity also represents a singular system and contains two second-class constraints. It would be interesting to quantize it in the BRST formalism (Becchi et al., 1976; Tyutin, 1975). This cannot be done directly; the model does not possess gauge invariances. It is necessary to implement some gauge invariances. This can be done by transforming the original second-class system into a first-class one in the original phase-space (Bizdadea and Saliu, 1995).

The forthcoming step is to quantize this model in the cosmological context using Feynman path integrals. With the new set of constraints, it is possible to choose appropriate gauge-fixing conditions and to recognize the gauge-invariant variables. Calculation of the corresponding path integrals is a work in progress and will be published elsewhere.

# DINAMIKA TAHIONSKOG POLJA SA VEZAMA U FLRW PROSTOR-VREMENU

*U ovom radu nastavljamo razmatranje tahionskog skalarnog polja opisanog Dirac-Born-Infeld (DBI) dejstvom sa vezama u kosmološkom kontekstu. Predloženo proširenje sistema, uvođenjem pomoćnog polja u minisuperprostornom modelu je detaljnije razmatrano. Novi ekvivalentni skup veza je postavljen tako da zadovoljava uobičajene uslove regularnosti.*

Ključne reči: *tahionsko polje, kosmologija, sistem sa vezama, Dirac-Bergmanov algoritam*